\documentstyle[12pt]{article}
\input epsf
\def\setb@se#1{\baselineskip=#1 \normalbaselineskip=#1}
\setb@se{14pt}
\newcommand{\be}{\begin{equation}}
\newcommand{\ee}{\end{equation}}

\begin{document}
\begin{titlepage}
\begin{flushright}
ETH-TH/97-16

ZU-TH/97-15

hep-th/9707176
\end{flushright}

\vspace{5 cm}

\begin{center}

{\huge Non-Abelian BPS Monopoles\\

\vspace{2 mm}

 in N=4 Gauged Supergravity}

\vspace{20 mm}

{\bf Ali H. Chamseddine }\footnote{e-mail: chams@itp.phys.ethz.ch}

\vspace{1 mm}

{\it Institute of Theoretical Physics, ETH-H\"onggenberg, 
CH--8093 Z\"urich, Switzerland}

\vspace{5 mm}

{\bf Mikhail S. Volkov}\footnote{e-mail: volkov@physik.unizh.ch}

\vspace{1 mm} 

{\it  Institute of Theoretical Physics, University of Z\"urich,
Winterthurerstrasse 190, CH--8057 Z\"urich, Switzerland}

\vspace{5 mm}

\end{center}

\noindent
We study static, spherically symmetric, and purely magnetic
solutions of SU(2) $\times$ SU(2) gauge supergravity in 
four dimensions. 
A systematic analysis of the supersymmetry conditions
reveals solutions 
which preserve 1/8 of the supersymmetries and are
characterized by a BPS-monopole-type
gauge field and a globally hyperbolic, everywhere regular geometry. 
These present the first known example 
of non-Abelian backgrounds in gauge supergravity
and in leading order effective string theory.

\vspace{30 mm}
\noindent
04.62.+v, 02.40.-k, 11.10.Ef, 11.25.-w, 11.30Pb
\end{titlepage}

\noindent
{\bf Introduction.--}
In the last few years there has been considerable interest in
supersymmetric solitons originating from
effective field theories of superstrings and heterotic strings
(see \cite{duff} for review).
These solutions play an important role in the study of the
non-perturbative sector of string theory and in understanding string
dualities. A characteristic feature of such solutions is that
supersymmetry is only partially broken, and associated
with each of the unbroken supersymmetries there is a Killing spinor
fulfilling a set of linear differential constraints.
The corresponding integrability conditions
can be formulated as a set of non-linear Bogomolny
equations for the solitonic background, which can often
be solved analytically.

The analysis of the supersymmetry conditions has proven to be
the efficient way of studying the non-perturbative sector.
So far, however, the investigations
have mainly been restricted to the Abelian theory, whereas
little is known about supergravity solitons
with non-Abelian gauge fields, which 
presumably is due to the complexity
of the problem. The known solutions
appear to be somewhat special, since they are
constructed either from the flat space configurations
of the Yang-Mills field,
by making use of the conformal invariance
(see \cite{duff}--\cite{monopole} and references therein),
or from the gravitating Abelian solitons, by identifying gravitational
and gauge connections \cite{renata}.
At the same time, the example of the well-known
(non-supersymmetric) Bartnik-McKinnon particles \cite{BK}
shows that the generic behavior of a gravitating Yang-Mills
field can be far more complex.

Motivated by this, we study
solitons in a four-dimensional supergravity model
with non-Abelian Yang-Mills multiplets.
The model
we consider is the N=4 gauged SU(2)$\times$SU(2) supergravity \cite{FS},
which can be regarded as N=1, d=10 supergravity compactified on
the group manifold S$^{3}\times$S$^{3}$.
The non-gauged version of the same model,
corresponding to the toroidal compactification of
ten-dimensional supergravity, has been
extensively studied in the past \cite{G}.
We investigate static, spherically
symmetric, purely magnetic field configurations and find in this case
analytically {\sl all} supersymmetric solutions.
Among them we discover
globally regular solutions characterized
by a BPS-monopole-type gauge field.
The corresponding geometry is globally hyperbolic and
does not belong to any standard type. 
It is worth noting that, although the Abelian solutions
in the model were studied long ago \cite{G},
to our knowledge, we present here the first example of
non-Abelian backgrounds.
At the same time, these are the first non-Abelian solutions 
of the {\sl leading order} equations of motion of the
effective string action. All other known solutions 
\cite{duff}--\cite{renata} have
gauge fields which originate from string corrections.

\noindent
{\bf The model.--}
The action of the N=4 gauged SU(2)$\times $SU(2) supergravity theory
includes a vierbein $e_{\mu }^{m}$, four Majorana spin-3/2 fields $\psi
_{\mu }\equiv \psi _{\mu }^{\rm{I}}$ $(\rm{I}=1,\ldots 4)$, vector
and pseudovector non-Abelian gauge fields $A_{\mu }^{a}$ and $B_{\mu }^{a}$
with independent gauge coupling constants $e_{A}$ and $e_{B}$, respectively,
four Majorana spin-1/2 fields $\chi \equiv \chi ^{\rm{I}}$, the axion
and the dilaton \cite{FS}. 
We consider the truncated theory specified by the
conditions $e_{B}=B_{\nu }^{a}=0$.  In addition, we require the vector
field $
A_{\mu }^{a}$ to be purely magnetic, which allows us to set the axion to
zero. After a suitable rescaling of the fields, the bosonic part of the
action reads
\begin{equation}
S=\int \left( -\frac{1}{4}\,R+\frac{1}{2}\,\partial _{\mu }\phi \,\partial
^{\mu }\phi -\frac{1}{4}\,e^{2\phi }\,F_{\mu \nu }^{a}F^{a\mu \nu }
+\frac{1}{8}\,e^{-2\phi }\right) \sqrt{-\bf{g}}\,d^{4}x,  \label{1}
\end{equation}
where $F_{\mu \nu }^{a}=\partial _{\mu }A_{\nu }^{a}
-\partial _{\nu }A_{\mu}^{a}+\varepsilon _{abc}A_{\mu }^{b}A_{\nu }^{c}$, 
and the dilaton potential can be viewed as an effective negative, 
position-dependent cosmological term $\Lambda(\phi)=-\frac{1}{4}e^{-2\phi}$.

For a purely bosonic
configuration, the supersymmetry transformation laws are \cite{FS}
\[
\delta \bar{\chi}=-\frac{i}{\sqrt{2}}\,\bar{\epsilon}\,\gamma ^{\mu
}\partial _{\mu }\phi -\frac{1}{2}e^{\phi }\,\bar{\epsilon}\,\alpha
^{a}F_{\mu \nu }^{a}\,\sigma ^{\mu \nu }+\frac{1}{4}\,e^{-\phi }\,\bar{
\epsilon},
\]
\begin{equation}
\delta \bar{\psi}_{\rho }=\bar{\epsilon}
\left( \overleftarrow{\partial}_{\rho 
}
-\frac{1}{2}\,\omega _{\rho mn}\,\sigma ^{mn}+\frac{1}{2}\,\alpha
^{a}A_{\rho }^{a}\right) -\frac{1}{2\sqrt{2}}\,e^{\phi }\,\bar{\epsilon}%
\,\alpha ^{a}F_{\mu \nu }^{a}\,\gamma _{\rho }\,
\sigma ^{\mu \nu }+\frac{i}{4
\sqrt{2}}\,e^{-\phi }\,\bar{\epsilon}\,\gamma _{\rho },  \label{2}
\end{equation}
the variations of the bosonic fields being zero. In these formulas, $
\epsilon \equiv \epsilon ^{\rm{I}}$ are four Majorana spinor
supersymmetry parameters, $\alpha ^{a}\equiv \alpha _{\rm{IJ}}^{a}$ are
the SU(2) gauge group generators, whose explicit form is given in \cite{FS},
and $\omega _{\rho mn}$ is the tetrad connection.

We shall consider static, spherically symmetric, purely magnetic
configurations of the bosonic fields,
and for this we parameterize the fields
as follows:
\[
ds^{2}=N\sigma ^{2}dt^{2}-\frac{dr^{2}}{N}-r^{2}(d\theta ^{2}+\sin
^{2}\theta \,d\varphi ^{2}),
\]
\begin{equation}
\alpha ^{a}A_{\mu }^{a}dx^{\mu }=w\ (-\alpha ^{2}\,d\theta +\alpha
^{1}\,\sin \theta \,d\varphi )+\alpha ^{3}\,\cos \theta \,d\varphi ,
\label{3}
\end{equation}
where $N$, $\sigma $, $w$, as well as the dilaton $\phi $, are
functions of the
radial coordinate $r$. The field equations, following from the action
(\ref{1}), read
\[
(rN)^{\prime }+r^{2}N\phi ^{\prime \,\,2}+U+r^2\Lambda(\phi)=1,
\]
\[
\left( \sigma Nr^{2}\phi ^{\prime }\right) ^{\prime
}=\sigma\, (U-r^2\Lambda(\phi)),
\]
\[
r^2\left( N\sigma e^{2\phi }\,w^{\prime }\right) ^{\prime }=\sigma e^{2\phi 
}\,
w(w^{2}-1),
\]
\begin{equation}
\sigma ^{\prime }=\sigma\, (r\phi ^{\prime \,\,2}+ 2e^{2\phi
}\,w^{\prime\, 2}/r),  \label{4}
\end{equation}
where $U=2e^{2\phi }\left( Nw^{\prime\, 2}
+(w^{2}-1)^{2}/2r^{2}\right)$.
Now, since we are unable to directly solve these equations, 
we shall consider the supersymmetry conditions for the fields (\ref{3}), 
which will give us a set of first integrals for the system (\ref{4}). 

\noindent
{\bf The supersymmetry conditions.--}
The field configuration (\ref{3}) is supersymmetric, provided that
there are non-trivial
supersymmetry Killing spinors $\epsilon$ for which the variations of the
fermion fields defined by Eqs. (\ref{2}) vanish. Inserting configuration
(\ref{3}) into Eqs. (\ref{2}) and putting $\delta\bar{\chi}=\delta\bar{\psi}
_{\mu}=0$, the supersymmetry constraints become a system of
equations for the four spinors $\epsilon^{\rm{I}}$.
The procedure which solves these equations is  rather
involved. For this reason we describe here only 
the principal steps of the  analysis.
First, since the background field is
static and spherically symmetric, we choose the spinors to be
time-independent and classify them with respect to the total angular
momentum $J=L+S+I$. Since spin $S$ and isospin $I$ are half-integer, $J$ is
integer, and we restrict to the $J=0$ sector. In this sector half of the 16
independent spinor components vanish (in the special representation chosen),
which in effect truncates half of the supersymmetries.
Hence, the supersymmetry
constraints $\delta \bar{\chi}=0$ and $\delta \bar{\psi}_{\mu }=0$ for $\mu
=t,\theta ,\varphi $ reduce to four systems of homogeneous algebraic
equations for the remaining eight spinor components, each system containing
eight equations, while the $\mu =r$ gravitino constraint becomes a system of
radial differential equations.

The consistency of the algebraic constraints requires that the determinants
of the corresponding coefficient matrices vanish and that the
matrices commute with each other. These consistency conditions
can be expressed by the  following relations for the background:
\begin{equation}
N\sigma ^{2}=e^{2(\phi -\phi _{0})},  \label{5a}
\end{equation}
\begin{equation}
N=\frac{1+w^{2}}{2}+e^{2\phi }\,\frac{(w^{2}-1)^{2}}{2r^{2}}+\frac{r^{2}}{8}
e^{-2\phi },  \label{5b}
\end{equation}
\begin{equation}
r\phi ^{\prime }=\frac{r^{2}}{8N}e^{-2\phi }\left( 1-4e^{4\phi }\,
\frac{(w^{2}-1)^{2}}{r^{4}}\right) ,  \label{5c}
\end{equation}
\begin{equation}
rw^{\prime }=-2w\frac{r^{2}}{8N}e^{-2\phi }\left( 1+2e^{2\phi }
\frac{w^{2}-1}{r^{2}}\right) ,  \label{5}
\end{equation}
with constant $\phi _{0}$. Under these conditions, the solution of the
algebraic constraints yields $\epsilon $ in terms of only two independent
functions of $r$. The remaining differential constraint then uniquely
specify these two functions up to two integration constants, which finally
corresponds to two unbroken supersymmetries. We therefore conclude that the
supersymmetry conditions for the bosonic background (\ref{3}) are given in
terms of Eqs. (\ref{5a})--(\ref{5}). Solutions of these Bogomolny equations
describe the BPS states in N=4 gauge supergravity with 1/8 of the
supersymmetries preserved (since the maximal possible number of
supersymmetries in the model is 16). One can verify that the Bogomolny
equations are compatible with the field equations (\ref{4}).

\noindent
{\bf The solution.--}
In order to find the general solution of the Bogomolny equations, we start
from the case where $w(r)$ is constant.
The only possibilities are $w(r)=\pm 1$ or $w(r)=0$.
For $w(r)=\pm 1$ the Yang-Mills field is a pure gauge,
and the equations imply that $\exp(-2\phi)=0$, which means that
$\phi(r)=\phi_0\rightarrow\infty$,
whereas the metric is flat. The $w(r)=0$ choice
corresponds to the Dirac monopole gauge field.
The general solution of the remaining non-trivial
Eq. (\ref{5c}) is then given by
$\phi+\ln(r/r_0)=r^2 e^{-2\phi}/4$,
with constant $r_0$; the corresponding metric 
turns out to be singular at the origin.

Suppose now that $w(r)$ is not a constant. Introducing the new variables $
x=w^{2}$ and $R^2=\frac{1}{2}r^{2}e^{-2\phi }$, Eqs. (\ref{5a})--(\ref{5})
become equivalent to one first order differential equation
\begin{equation}
2xR\, (R^2+x-1)\frac{dR}{dx}+(x+1)\,R^2+(x-1)^{2}=0.  \label{6}
\end{equation}
If $R(x)$ is known,
the radial dependence of the functions, $x(r)$
and $R(r)$, can be determined from
(\ref{5c}) or (\ref{5}). Eq. (\ref{6}) is
solved by the following substitution:
\begin{equation}
x=\rho^{2}\,e^{\xi (\rho)},\ \ \ \ \ \ \
R^2=-\rho\frac{d\xi (\rho)}{d\rho}-\rho^{2}\,e^{\xi
(\rho)}-1,  \label{7}
\end{equation}
where $\xi (\rho)$ is obtained from
\begin{equation}
\frac{d^{2}\xi (\rho)}{d\rho^{2}}=2\, e^{\xi (\rho)}.  \label{8}
\end{equation}
The most general (up to reparametrizations)
solution of this equation which ensures that $R^2>0$
is $\xi (\rho)=-2\ln\sinh(\rho-\rho_0)$.
This gives us the general solution of  Eqs. (\ref{5a})--(\ref{5}). 
The metric is non-singular at the origin if only 
$ \rho_{0}=0 $, in which case
\begin{equation}                    \label{R}
R^{2}(\rho)=2\rho\coth \rho-\frac{\rho^{2}}{\sinh ^{2}\rho}-1,
\end{equation}
one has $R^{2}(\rho)=\rho^{2}+O(\rho^{4})$ as $\rho\rightarrow 0$,
and $R^{2}(\rho)=2\rho+O(1)$ as $\rho\rightarrow \infty $.
The last step is to obtain $r(s)$ from Eq. (\ref{5}), which finally
gives us a family of completely regular solutions
of the Bogomolny equations:
\begin{equation}
d{ s}^{2}=a^2\, 
\frac{\sinh \rho}{R(\rho)}\left\{
dt^{2}-d\rho^{2}-R^{2}(\rho)(d\vartheta ^{2}+\sin ^{2}\vartheta d\varphi
^{2})\right\} ,  \label{10}
\end{equation}
\begin{equation}
w=\pm \frac{\rho}{\sinh \rho},\ \ \ \
e^{2\phi }=a^2\, \frac{\sinh \rho}{2\,R(\rho)},
\label{11}
\end{equation}
\begin{figure}
\epsfxsize=10cm
\centerline{\epsffile{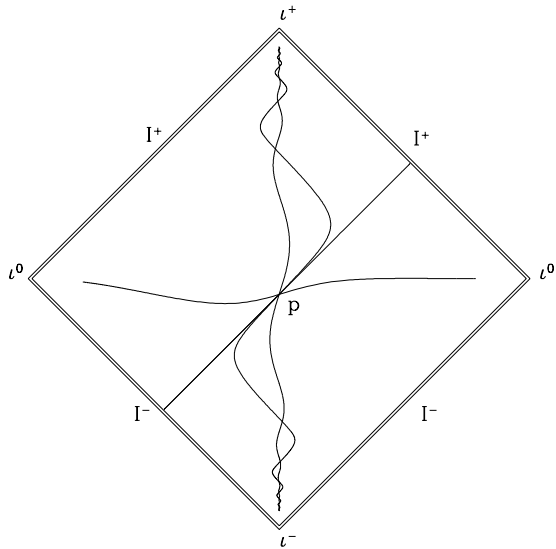}}
\caption{The conformal diagram for the spacetime described by the
line element (13).}
\end{figure}

\noindent
where $0\leq \rho<\infty $, $R(\rho)$ is given by Eq. (\ref{R}), 
and we have chosen in Eq. (\ref{5a})
$2\phi_0=-\ln 2$. The appearance of the free parameter 
$a$ in the solution reflects the scaling symmetry of
Eqs. (\ref{5a})--(\ref{5}): $r\rightarrow ar$, 
$\phi\rightarrow\phi+\ln a$. 
The geometry described by the line element (\ref{10}) is
everywhere regular,
the coordinates covering the whole space whose
topology is R$^4$. The geometry becomes flat at the
origin, but asymptotically it is not flat,
even though the cosmological term $\Lambda(\phi)$ vanishes at infinity.
We thus can not assign a total energy to the solution. 
Specifically, in the asymptotic region
all curvature invariants tend to zero, however, not fast enough.
The Schwarzschild metric functions for $r\rightarrow\infty$ are 
$N\propto \ln r$ and $N\sigma^2\propto r^2/4\ln r$,
the non-vanishing Weyl tensor invariant being $\Psi_2\propto-1/6r^2$.

The global structure of the solution is well illustrated by the
conformal diagram. Inspecting the $t$--$\rho$ part of the metric,
it is not difficult to see that the conformal
diagram in this case is actually identical to the one for
Minkowski space, even though the geometry is not asymptotically flat
(see Fig.1). The spacetime is therefore geodesically complete and
globally hyperbolic. The latter property is quite remarkable, since
global hyperbolicity is usually lacking for
the known supersymmetry backgrounds in gauged supergravity models.
The geodesics through a spacetime point $p$
are shown in the diagram, each geodesic approaching infinity for
large absolute values of the affine parameter.
Although the global behavior of geodesics  is
similar to that for Minkowski space,
they locally behave differently.
For $\rho<\infty$ the cosmological term $\Lambda(\phi)$ is non-zero
and negative, thus having the focusing effect on timelike geodesics,
which makes them oscillate around the origin.
Unlike the situation in the anti-de Sitter case, each geodesic
has its own period of oscillations,
such that the geodesics from a point $p$
never refocus again.  

The shape of the gauge field amplitude $w(\rho)$, given by Eq. (\ref{11}),
corresponds to the gauge field of
the regular magnetic monopole type.
In fact, replacing $\rho$ by $r$, the amplitude
exactly coincides with that for the flat space BPS  solution.
This result is quite surprising, since the model has no
Higgs field, in which case 
it would be natural to expect the existence of only
neutral solutions \cite{BK}.
Note that all known stringy monopoles in four dimensions
\cite{duff}, \cite{monopole} contain a Higgs field.  

In conclusion, Eqs. (\ref{10}), (\ref{11}) describe 
globally regular, supersymmetric backgrounds of a new type.
The existence of unbroken supersymmetries suggests
that the configurations should be stable,
and we expect that the stability
proof can be given along the same lines as in \cite{AD}.
Being solutions of N=4 quantum supergravity in four dimensions,
they presumably
receive no quantum corrections. On the other hand,
they can be considered in the framework of 
string theory, and then the issue of 
string corrections can be addressed.
In order to study this problem,
we first of all need to lift the solutions to ten dimensions.
Although the process is rather involved, one can show
that such a lifting is indeed possible.

\noindent
{\bf Acknowledgments.--} MSV thanks M. Heusler and N. Straumann
 for the reading of the manuscript, and acknowledges 
the support of the Swiss National Science Foundation.

\newpage

\end{document}